\documentclass[twocolumn,trackchanges]{aastex701}

\usepackage{amsmath}

\graphicspath{{./}{fig/}}

\journalinfo{Publications of the Astronomical Society of the Pacific, in press} 

\received{2025 November 20}
\revised{2025 December 15}
\accepted{2025 December 17}

\begin{document}

\title{Rapid Brightening of 3I/ATLAS Ahead of Perihelion}

\shorttitle{Rapid Brightening of 3I/ATLAS Ahead of Perihelion}
\shortauthors{Q. Zhang \& K. Battams}

\correspondingauthor{Qicheng Zhang}

\author[0000-0002-6702-191X]{Qicheng Zhang}
\affiliation{Lowell Observatory, 1400 West Mars Hill Road, Flagstaff, AZ 86001, USA}
\email[show]{qicheng@cometary.org}

\author[0000-0002-8692-6925]{Karl Battams}
\affiliation{US Naval Research Laboratory, 4555 Overlook Avenue, SW, Washington, DC 20375, USA}
\email{karl.battams.civ@us.navy.mil}

\begin{abstract}
Interstellar comet 3I/ATLAS has been approaching its 2025 October 29 perihelion while opposite the Sun from Earth, hindering ground-based optical observations over the preceding month. However, this geometry placed the comet within the fields of view of several space-based solar coronagraphs and heliospheric imagers, enabling its continued observation during its final approach toward perihelion. We report photometry from STEREO-A's SECCHI HI1 and COR2, SOHO's LASCO C3, and GOES-19's CCOR-1 instruments in 2025 September--October, which show a rapid rise in the comet's brightness scaling with heliocentric distance $r$ as $r^{-7.5\pm1.0}$. CCOR-1 also resolves the comet as an extended source with an apparent coma ${\sim}4'$ in diameter. Furthermore, LASCO/CCOR-1 color photometry shows the comet to be distinctly bluer than the Sun, consistent with gas emission contributing a substantial fraction of the visible brightness near perihelion.
\end{abstract}

\keywords{Comets --- Broad band photometry --- Coronagraphic imaging}

\section{Introduction}

Interstellar comet 3I/ATLAS was discovered on 2025 July 1 at a heliocentric distance $r=4.5$~au while inbound toward its 2025 October 29 perihelion at $r=1.36$~au \citep{denneau2025}. At the time, it was comparable in brightness at similar $r$ to the previous interstellar comet, 2I/Borisov \citep{seligman2025}. However, that similarity turned out to be coincidence of timing, as 2I brightened slowly with decreasing heliocentric distance $r$ on its approach, with (observer distance $\varDelta$-corrected) brightness scaling as $r^{-n}$ with only $n\approx2$---close to that of an inert, reflecting object---from the earliest pre-discovery observations at $r\approx8$~au to its perihelion at $r\approx2$~au. Meanwhile, 3I varied in its rate of brightening, but exhibited a much steeper overall $n=3.8\pm0.3$ from its earliest pre-discovery observations at $r\approx6$ through the pre-perihelion observations at $r\approx2$~au \citep{ye2025,tonry2025,jewitt2025}, perhaps reflecting a difference in nucleus properties or the faster speed at which it approached the Sun.

These photometric observations before and soon after discovery largely measured light scattered by 3I's dust, given the absence of gas emission in near-discovery optical spectra \citep[e.g.,][]{opitom2025,kareta2025}. However, continued observations from $r=4$ to 2~au revealed the production of several gaseous species---including OH, CN, \ion{Ni}{1}, and \ion{Fe}{1}---to have been rising far more steeply, with $n>8$ \citep{rahatgaonkar2025,hutsemekers2025}. By 2025 September ($r\sim2$~au), the comet had developed a green halo enveloping the existing dust coma and tail resembling a typical C$_2$ gas coma (e.g., images in BAA comet image archive\footnote{\url{https://britastro.org/cometobs/3i}}). Magnitudes reported to the COBS database \citep{zakrajsek2018}\footnote{\url{https://cobs.si}} also began deviating sharply above the prior trend, perhaps reflecting the surging gas emission.

Optical observations, however, rapidly became difficult from the ground at $r\lesssim2$~au as the comet approached its 2025 October 21 superior conjunction at only $1^\circ\llap{.}9$ solar elongation from Earth, leaving a gap in this dataset just as the comet approaches its maximum heating at perihelion. However, the comet remained observable during this period to a number of spacecraft around the solar system that are not designed for comet observations, but can nonetheless help to fill in this gap \citep{eubanks2025}. Among these spacecraft are several solar observatories with cameras that continually monitor the corona and inner heliosphere, but which also routinely record the passage of comets that happen to cross their fields of view \citep{battams2017}. Here, we present observations from solar coronagraphic and heliospheric imagers onboard the STEREO-A, SOHO, and GOES-19 spacecraft, as well as a cursory analysis of their implications for the 3I's activity on its final approach toward perihelion.

\section{Observations}

\subsection{Instruments}

We present observations of 3I from four cameras onboard three space-based solar observatories, as summarized in Table~\ref{tab:obs} and detailed below:

\begin{deluxetable*}{llclcccc}
\tablecaption{Summary of observations of 3I/ATLAS}
\label{tab:obs}

\tablecolumns{7}
\tablehead{
\colhead{Spacecraft} & \colhead{Instrument} & \colhead{Bandpass (nm)\tablenotemark{a}} & \colhead{Observation Dates\tablenotemark{b}} & \colhead{$r$ (au)\tablenotemark{c}} & \colhead{$\varDelta$ (au)\tablenotemark{d}} & \colhead{$\alpha$ ($^\circ$)\tablenotemark{e}} & \colhead{$\varepsilon$ ($^\circ$)\tablenotemark{f}}
}

\startdata
STEREO-A & SECCHI HI1 & 595--720 & 2025 Sep 11--26 & 1.79--2.20 & 2.75--3.04 & 2.7--10.4 & 5.0--24.3 \\
 & SECCHI COR2 & 670--750 & 2025 Sep 28--Oct 2 & 1.68--1.77 & 2.64--2.73 & 1.7--2.4 & 3.1--4.2 \\
SOHO & LASCO C3 & 520--770\tablenotemark{g} & 2025 Oct 15--26 & 1.36--1.45 & 2.33--2.42 & 1.8--5.9 & 2.6--8.4 \\
GOES-19 & CCOR-1 & 470--740 & 2025 Oct 18--24 & 1.37--1.42 & 2.35--2.41 & 1.9--3.7 & 2.6--5.3
\enddata

\tablenotetext{a}{FWHM wavelength range \citep{halain2012,jones2018,thernisien2025}.}
\tablenotetext{b}{Dates on which the comet was within the instrument field of view, even if unobservable due to interference from stray light, background objects, coronal structure, etc.}
\tablenotetext{c}{Comet heliocentric distance.}
\tablenotetext{d}{Observer--comet distance.}
\tablenotetext{e}{Phase angle.}
\tablenotetext{f}{Solar elongation.}
\tablenotetext{g}{For the primary, Clear filter; measured from published transmission + quantum efficiency curves.}
\end{deluxetable*}

\begin{enumerate}
\item STEREO \citep[``Solar TErrestrial RElations Observatory'';][]{kaiser2008} is comprised of two effectively identical spacecraft launched in 2006: STEREO-A, which orbits slightly interior to/more quickly than Earth, and STEREO-B, which orbits slightly exterior to/more slowly than Earth. STEREO-B malfunctioned in, and has been defunct since, 2014 \citep{ossing2018}, so only STEREO-A observed 3I. The spacecraft carries a SECCHI \citep[``Sun Earth Connection Coronal and Heliospheric Investigation'';][]{howard2008} instrument suite, of which we report observations from two cameras:
\begin{enumerate}
\item HI1 \citep[``Heliospheric Imager 1'';][]{eyles2009} observes a $20^\circ\times20^\circ$ field at a normally $2\times2$-binned scale of 72~arcsec~px$^{-1}$, centered $14^\circ$ from the Sun along the Sun--Earth line. The camera observes through a filter with a pre-flight full-width half-maximum (FWHM) wavelength span of $\sim$615--740~nm, as well as a blue leak near 400~nm and a red leak near 1000~nm \citep{bewsher2010}, with later testing with a flight spare suggesting that the bandpass may have shifted blueward by 15--20~nm \citep{halain2012}. STEREO-A HI1 observed 3I over 2025 September 11--27.
\item COR2 (``CORonagraph 2'') covers ${\sim}0^\circ\llap{.}7$--$4^\circ$ elongation at 15~arcsec~px$^{-1}$ through a $\sim$670--750~nm bandpass filter as well as a rotating polarizer. It collects two types of images: polarized sequences of three images with the polarizer rotated $0^\circ$, $120^\circ$, and $240^\circ$, and non-polarized images made from a sum of consecutive exposures with the polarizer at $0^\circ$ and $90^\circ$. Most of the data is of the latter variety, and we only present results from these unpolarized data. STEREO-A COR2 observed 3I over 2025 September 28--October 2. Note that the comet's superior conjunction from STEREO-A was on September 30---well before the October 21 conjunction from Earth---due to STEREO-A orbiting ${\sim}47^\circ$ ahead of Earth during this time.
\end{enumerate}
\item SOHO \citep[``SOlar and Heliospheric Observatory'';][]{domingo1995}, launched 1995, orbits the Sun--Earth L1 point, carrying the LASCO \citep[``Large Angle and Spectrometric COronagraph'';][]{brueckner1995} coronagraphs. Its wide field coronagraph, C3, covers ${\sim}1^\circ$--$8^\circ$ elongation at 56~arcsec~px$^{-1}$ primarily through its $\sim$520--770~nm Clear filter, although it also has an assortment of less frequently used color filters. LASCO C3 observed 3I over 2025 October 15--26, shortly before the comet's October 29 perihelion.
\item GOES-19 was launched in 2024, and is primarily a weather satellite operating in a geostationary orbit. However, it also carries the CCOR-1 \citep[``Compact CORonagraph 1'';][]{thernisien2025} coronagraph for operational space weather monitoring. CCOR-1 covers ${\sim}1^\circ$--$6^\circ$ elongation at 19~arcsec~px$^{-1}$ with a $\sim$470--740~nm bandpass, and observed 3I concurrently with LASCO C3 over 2025 October 18--24.
\end{enumerate}

All observations of 3I by these instruments were made as part of their standard observing routine.

\subsection{Data}

We processed images from all instruments similarly following a procedure similar to that used by \citet{zhang2023}, and started from the level-2 HI1\footnote{Available online at \url{https://stereo-ssc.nascom.nasa.gov/data/ins_data/secchi/secchi_hi/L2/a/img/hi_1/20250926/} for 2025 September 26 data; replace the 20250926 in the URL with the date of interest in YYYYMMDD format.}, level-0 COR2\footnote{Available online at \url{https://stereo-ssc.nascom.nasa.gov/data/ins_data/secchi/L0/a/img/cor2/20251001/} for 2025 October 1 data; likewise, replace the 20251001 with the date of interest}, level-0.5 C3\footnote{Available online at \url{https://lasco-www.nrl.navy.mil/lz/level_05/251020/c3/} for 2025 October 20 data; replace the 251020 with the date of interest in YYMMDD format.} (plus level-1 vignetting/bias correction), and level-1a CCOR-1 data\footnote{Available online at \url{https://noaa-nesdis-swfo-ccor-1-pds.s3.amazonaws.com/index.html\#SWFO/GOES-19/CCOR-1/ccor1-l1a/}.}. We first derived astrometric solutions for all coronagraph (COR2, C3, and CCOR-1) frames using Gaia DR3 \citep{gaia2023}, but used the existing HI1 solutions. We subtracted a stray light/corona model from all coronagraph frames, derived from an average of all frames in each dataset; level-2 HI1 data already includes this correction. We also subtracted a stellar background from each HI1 frame, derived from frames several hours away to avoid self-subtraction of the moving comet. 

For all frames, we extracted cutouts centered on the ephemeris position (JPL orbit \#27) and stacked these comet-centered cutout frames. Stacking was critical to our analysis to smooth out background features and single frame image defects (e.g., stars, coronal structure, cosmic rays/solar energetic particles, etc.) and to obtain images with sufficiently high S/N for analysis. In fact, only in CCOR-1 data is 3I only clearly visible in individual frames; in HI1, the comet can be marginally seen in a subset of frames at the noise level, while in COR2 and C3, the comet cannot be seen at all without stacking. No offset in the comet's observed position could be distinguished from the ephemeris position at the resolution of any of the data, so we applied no further corrections to the ephemeris position.

\begin{figure}
\centering
\includegraphics[width=0.59\linewidth]{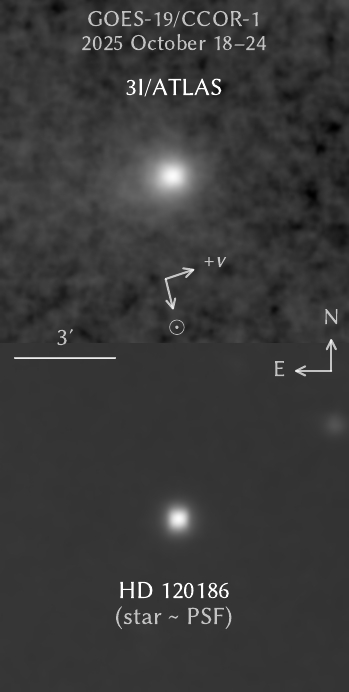}
\includegraphics[width=0.385\linewidth]{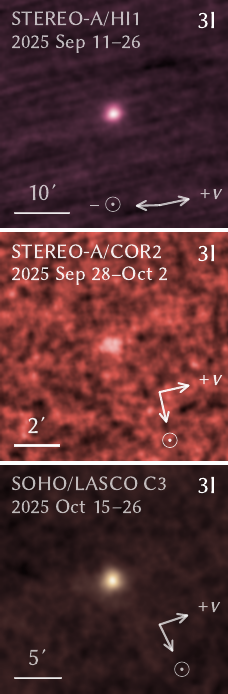}
\caption{Left: Stack of all CCOR-1 frames of 3I/ATLAS (top), and an equivalent stack centered on a nearby star on the same frames, approximating the PSF (bottom). Right: Similar stacks of all HI1 (top), COR2 (middle), and LASCO C3 Clear (bottom) frames of 3I. All stacks are aligned with north up. The heliocentric velocity ($+v$), and sunward ($\odot$) or antisunward ($-\odot$) directions are labeled for the comet at the midpoint time.}
\label{fig:img}
\end{figure}

Figure~\ref{fig:img} shows the stack on the comet through each of the four instruments. The CCOR-1 stack resolves the comet as extended with a ${\sim}4'$ diameter coma, compared to a similarly stacked star nearby that serves as an approximate point spread function (PSF). No tail is clearly visible. Note, however, that while an ion tail nominally points in antisunward direction with a ${\sim}10^\circ$ aberration lag at the comet's $\sim$70~km~s$^{-1}$ near perihelion through a typical $\sim$400~km~s$^{-1}$ solar wind, it can vary from this direction by another ${\sim}10^\circ$ with directional deviations in the solar wind \citep{fernandez1997}. The long duration of the stack combined with the low (${<}4^\circ$) phase angle means that such an ion tail could theoretically point in any projected direction on the individual frames, and thus, may simply be smeared out in the stack, perhaps contributing to the apparent coma. In addition, any physically antisunward dust tail would also be highly foreshortened by this geometry. Note also that the sunward direction rotated by $125^\circ$ over the time range of the stack, so any angularly sunward-aligned structures would likewise be smeared out; however, the sunward-aligned stack does not appear qualitatively different from the presented north-aligned stack, so is not shown.

For time series photometry, we stacked the cutouts over over 1~day windows, except for COR2 where we split the data into a pair of 2.5~day/1.7~day stacks with similar S/N, due to the low total S/N of that dataset being insufficient for useful 1~day stacks. We measured the flux within radii of $3'$ for HI1, $1'$ for COR2, and $2'\llap{.}5$ for both C3 and CCOR-1. These apertures far exceed the ${<}1'$ dust coma/tail reported by \citet{jewitt2025} in ground-based observations contemporaneous with the HI1, and are also comparable to or larger than the Haser scale lengths for most optical gases each bandpass is sensitive to \citep{cochran1992}, thus capturing nearly all of the comet's expected dust and gas flux. We verified this conclusion by performing photometry on the full image stack for each dataset (which has higher S/N than individual time series stacks, permitting photometry with larger apertures) with both the adopted apertures and ones 2$\times$ the radii, and found the former captured $\gtrsim$90\% of the flux within the latter.

We then converted the instrumental fluxes to the conventional solar magnitude system (i.e., Sun as $-26.76$ at $r=1$~au, its Johnson $V$ magnitude; \citealt{willmer2018}), which is convenient for distinguishing differences from solar color. We used existing time-dependent photometric calibrations of \citet{tappin2022} for HI1 and \citet{zhang2023} for C3, extrapolated to the appropriate epochs in 2025. For COR2 and CCOR-1, we followed the same procedure \citet{zhang2023} used for LASCO, and derived zero-point magnitudes of $12.5\pm0.1$ and $14.7\pm0.1$ from 2025 April and May observations of the solar analog star 39~Tau \citep[$V$ magnitude 5.9, differing from solar color by $<$0.1~mag over the relevant wavelengths;][]{farnham2000}, respectively.

\begin{figure}
\centering
\includegraphics[width=\linewidth]{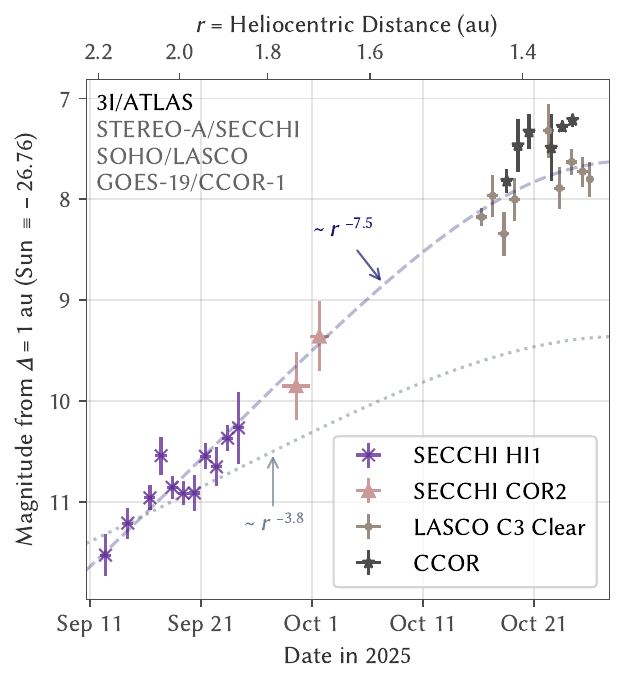}
\caption{Light curve of daily (and COR2 2.5~day/1.7~day) photometry, corrected to $\varDelta=1$~au, along with the $r^{-7.5}$ best fit brightness scaling and an $r^{-3.8}$ curve matching the previously reported trend at $r\gtrsim2$~au with an arbitrary vertical placement. Note that the $r^{-7.5}$ curve is specifically scaled for the C3 Clear magnitude of the comet, to which the CCOR-1 points can be converted with a $\sim$0.4~mag shift (i.e., the CCOR-1--Clear color from Table~\ref{tab:color}), and from which we estimate the HI1/COR2 points may likewise be plausibly offset $\pm0.4$~mag. Note also that the magnitude error bars indicate only the flux measurement uncertainty (i.e., the noise in the S/N of each detection) and do not include the estimated $\lesssim$0.1~mag uncertainties in the photometric calibrations.}
\label{fig:lc}
\end{figure}

Figure~\ref{fig:lc} presents all of the resulting photometry that detected the comet at $>3\sigma$, corrected to a common $\varDelta=1$~au. No phase correction has been applied, as the data cover a narrow phase angle range $\alpha=1^\circ\llap{.}7$--$10^\circ\llap{.}4$ corresponding to only 0.3~mag variation in dust brightness under the standard Schleicher--Marcus phase function \citep{schleicher2011,marcus2007}, which is further diluted by an unknown dust-to-gas brightness ratio. Moreover, the phase function of even the dust alone may not necessarily match that of solar system comets given the unusual polarization of sunlight scattered by 3I's dust that is indicative of unusual grain properties and scattering behavior \citep{gray2025}.

Note, however, that any direct photometric fit across all points would implicitly assume the comet is solar-colored between the relevant instrument bandpasses, which is not necessarily true, as demonstrated by the offset between the contemporaneous CCOR-1 and LASCO C3 Clear photometry. While all four bandpasses have similar effective wavelengths of $\sim$600--700~nm, the variations in their spectral profiles still produce differences in sensitivity to common optical gas species, particularly C$_2$, NH$_2$, and CN \citep{ye2014,jones2018}.

LASCO actually has the capability to directly measure color: while it takes the vast majority of its data through its Clear filter (including the data used for the time series photometry in Figure~\ref{fig:lc}), it also presently takes one full resolution frame per day through each of its Blue, Orange, DeepRd (``deep red''), and IR (``infrared'') color filters, whose properties are detailed in \citet{brueckner1995} and \citet{thernisien2006}. Stacking all the frames through each color filter yields detections of 3I in only the Blue and Orange stacks. The corresponding photometry yields a Blue--Clear color of ($-0.7\pm0.3$)~mag and an Orange--Clear color of ($-0.4\pm0.2$)~mag---distinctly bluer than solar color, in contrast to the comet's red dust \citep{opitom2025,kareta2025}. The Blue bandpass efficiently transmits the C$_2$ Swan bands, while the Orange bandpass is less sensitive to C$_2$ but efficiently transmits several NH$_2$ bands \citep{kawakita2002,jones2018}, suggesting that the presence of these species could be at least partly responsible for the observed deviations from solar colors. The LASCO C3 photometry can also be compared with the contemporaneous CCOR-1 photometry, which are measured with the same photometric aperture size from nearly the same location (GOES-19, in geosynchronous orbit, was only $\sim$0.4\% farther from the comet than SOHO, in its L1 halo orbit). CCOR-1's bandpass is similar to, but $\sim$40~nm bluer than, the C3 Clear bandpass. The difference puts the strongest, $\Delta\nu=0$ band of C$_2$---which falls just blueward of C3 Clear's FWHM interval---cleanly within CCOR-1's bandpass, substantially elevating its sensitivity to C$_2$ \citep{sivaraman1988}. Indeed, the comet's CCOR-1 brightness falls between the C3 Blue and Clear values, with a CCOR-1--Clear color of ($-0.4\pm0.2$)~mag. Table~\ref{tab:color} provides the observed (i.e., not $\varDelta$-corrected) magnitudes through each filter from which these colors were derived.

\begin{deluxetable}{lcc}
\tablecaption{LASCO C3 + CCOR-1 color photometry}
\label{tab:color}

\tablecolumns{3}
\tablehead{
\colhead{Filter} & \colhead{Bandpass (nm)\tablenotemark{a}} & \colhead{Observed Magnitude\tablenotemark{b}}
}

\startdata
C3 Clear & 520--770 & $9.82\pm0.07$ \\
C3 Blue & 440--520 & $9.13\pm0.20$ \\
C3 Orange & 550--630 & $9.42\pm0.14$ \\
C3 DeepRd & 720--800 & $>$9.3 ($3\sigma$) \\
C3 IR & 840--900 & $>$9.3 ($3\sigma$) \\
CCOR-1 & 470--740 & $9.40\pm0.08$
\enddata

\tablenotetext{a}{FWHM wavelength range; computed for LASCO C3 from published filter transmission + quantum efficiency curves; CCOR-1 values from \citet{thernisien2025}.}
\tablenotetext{b}{Comet magnitude measured from a stack of all frames through each filter. Note that the stated uncertainties do not include the estimated $\lesssim$0.1~mag uncertainties in the photometric calibrations.}
\end{deluxetable}

\begin{figure*}
\includegraphics[width=0.95\linewidth]{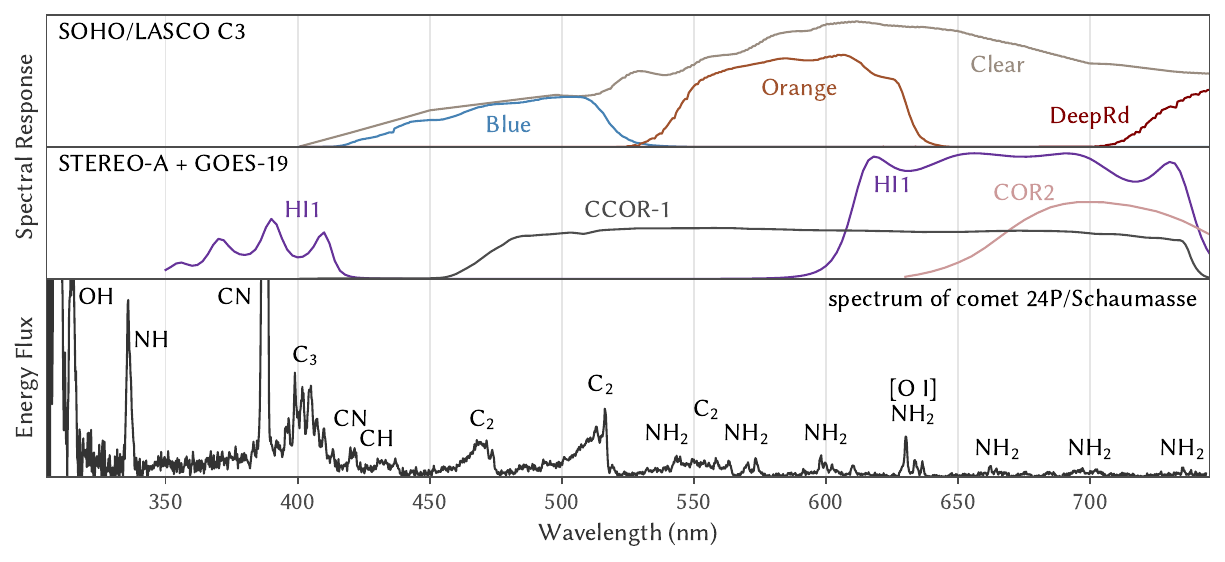}
\caption{The spectral responses corresponding to the presented observations (upper two panels) compared with a continuum-subtracted emission spectrum of a typical solar system comet, 24P/Schaumasse (bottom panel; courtesy of Q. Ye and C.~A. Schmidt), with prominent emission features labeled \citep{cochran1992,brown1996,farnham2000}. All panels have arbitrary vertical scaling. While several bandpasses extend past the red end of this spectrum, only a few minor NH$_2$ and CN bands are present there; those are typically dwarfed by the dust continuum, so are not expected to substantially contribute to the observed broadband flux. We caution, however, that this spectrum (1) is not of 3I itself, which may differ substantially in its gas abundances, and (2) only samples emission from a small fraction of the coma, and the relative brightness of different species may not closely match that over a much larger fraction of the coma (i.e., as captured by our large aperture photometry) given the wide variation in scale lengths between species \citep{ahearn1995}.}
\label{fig:bands}
\end{figure*}

Unlike C3 Clear and CCOR-1, neither HI1 nor COR2 are sensitive to the C$_2$ Swan bands. However, both instruments have much narrower primary bandpasses covering multiple NH$_2$ bands, offsetting the lack of C$_2$ sensitivity with increased NH$_2$ sensitivity. While these NH$_2$ bands are normally much fainter than the C$_2$ Swan bands \citep{cochran1992}, 3I has been found to be C$_2$-depleted \citep{hutsemekers2025}, which may improve the trade balance if C$_2$ is still depleted. HI1's blue leak also provides slight sensitivity to CN and C$_3$, while it remains unknown if COR2 has similar leaks. Figure~\ref{fig:bands} provides a visualization of these bandpasses relative to these and optical gas emission features of a typical solar system comet. A rigorous quantitative analysis requires an accurate inventory of the comet's optical emission and dust brightness, neither of which is available.

For our cursory analysis, we consider that the comet's brightness in HI1 and COR2 would likely be more similar to that in C3 Clear than CCOR-1 given C3 Clear's much lower relative C$_2$ sensitivity, and therefore convert the CCOR-1 photometry to provide the equivalent C3 Clear brightness (i.e., by applying a +0.4~mag offset) for the global fit. Given that the $\Delta\nu=0$ C$_2$ Swan band largely captured by CCOR-1, but excluded by C3 Clear, comprises most of the brightness of C$_2$ \citep{sivaraman1988}, that C$_2$ sensitivity is the primary difference between C3 Clear and HI1/COR2, and that any gas emission in HI1/COR2 is either also within the C3 Clear bandpass or only weakly transmitted in a red or blue leak, we expect the offset between HI1/COR2 and C3 Clear will likely be no more than the CCOR-1--Clear difference. Therefore, we incorporate into our global fit a potential offset of HI1/COR2 brightness from C3 Clear by $\pm$0.4~mag using Monte Carlo sampling to fit the light curve with a corresponding ensemble of HI1/COR2 offsets. With these color allowances, we obtain $n=7.5\pm1.0$ (with a corresponding normalized C3 Clear magnitude $M_1=5.1\pm0.2$ at $r=\varDelta=1$~au)---much steeper than the $n=3.8\pm0.3$ previously reported for the comet's earlier brightening trend at $r\gtrsim2$~au \citep{jewitt2025}.

Note that the standard Johnson $V$ bandpass is far more sensitive to C$_2$ and NH$_2$ than the C3 Clear bandpass due to the former covering a much narrower range isolating this emission, like C3 Blue or Orange. Consequently, we expect the C3 Blue or Orange brightness---$\sim$0.4--0.7~mag above the fitted curve---to better represent the true $V$ brightness of the comet.

\section{Discussion}

While ground-based optical observations have been hindered by the comet's low solar elongation over much of our observation period, radio observations are less adversely impacted by this geometry, and \citet{crovisier2025} recently reported a detection of OH radio emission over October 13--19 ($r=1.4$~au) corresponding to a production rate of $(5.7\pm0.6)\times10^{28}$~molecules~s$^{-1}$. For comparison, the last pre-conjunction optical OH measurement by \citet{hutsemekers2025} on September 12 ($r=2.19$~au) yielded a production rate of $(1.4\pm0.3)\times10^{27}$~molecules~s$^{-1}$. The power curve linking these two production rate measurements of OH---often used as a proxy for H$_2$O, which is typically the primary source of OH for comets---has $n=8.3\pm0.6$, which is quite similar to the $n=7.5\pm1.0$ we obtained for the comet's optical brightness.

Note that these two $n$ are not necessarily directly comparable as the optical brightness of a comet does not scale with its H$_2$O or overall gas production unless all of the following conditions are met:

\begin{enumerate}
\item The observed optical brightness is predominantly from gas emission and not dust, whose brightness involves a further $r^{-2}$ term corresponding to the variation of intercepted sunlight, as well as complications related to the dust's residence time in the photometric aperture.
\item The aperture encompasses all of the emission from the gas from formation/release through photodissociation or ionization, in which case, measured brightness has a $r^2$ scaling from gas lifetime that exactly offsets the $r^{-2}$ scaling of fluorescence efficiency.
\item The Swings effect \citep{swings1941}---variations in the fluorescence efficiency with heliocentric radial velocity---is negligible.
\item The mixing ratios of the optical gases is constant over the observation period, so the production rates of these gases scale with the H$_2$O or overall gas production rate.
\end{enumerate}

Condition 1 is plausible given the comet's lack of a prominent dust tail and blue color, but a dominant dust contribution to the brightness remains compatible with the presented data---particularly for HI1 and COR2, which are insensitive to C$_2$. NH$_2$ is the major gas species that HI1 and C3 Clear (and CCOR-1, corrected to C3 Clear)---which anchor the two ends of the fit---are both sensitive to. Condition 2 is satisfied for NH$_2$, as its Haser scale lengths \citep{fink1991} are contained within all the aperture radii we used, apart from the low S/N COR2 photometry that negligibly contributes to the fit. The Swings effect is usually neglected for NH$_2$ when deriving their production rates \citep{cochran1992}, satisfying condition 3. The status of condition 4 is not directly known, although the closeness of the $n$ in the comparison above and the possible satisfaction of the other three conditions hints at its possible validity.

Following its 2025 October 29 perihelion, 3I makes a return to twilight and subsequently dark, night skies over 2025 November--December. Ground-based observations will then, once again, be able to characterize the comet in far greater detail than possible with the data we have presented, whose value lies primarily in bridging the gap in ground-based optical observations during a critical period in the comet's evolution. Our cursory analysis of this data indicates the comet will likely emerge from conjunction considerably brighter than when it entered, with an extrapolated geocentric $V$ magnitude of $\sim$9 at perihelion, perhaps driven by prominent, visible gas emission.

The reason for 3I's rapid brightening, which far exceeds the brightening rate of most Oort cloud comets at similar $r$ \citep{holt2024}, remains unclear. It is possible that its H$_2$O sublimation had been held down earlier by cooling from its CO$_2$ sublimation, which remained unusually dominant at $r\sim3$~au \citep{lisse2025,cordiner2025}, perhaps related to its rapid approach toward the Sun compared to other comets. Oddities in nucleus properties like composition, shape, or structure---which might have been acquired from its host system or over its long interstellar journey---may likewise contribute. Without an established physical explanation, the outlook for 3I's post-perihelion behavior remains uncertain, and a plateau in brightness---or even a brief continuation of its pre-perihelion brightening---appears as plausible as rapid fading past perihelion. Continued observations may help provide a more definitive explanation for the comet's behavior.

\section{Conclusions}

We presented observations of interstellar comet 3I/ATLAS in solar coronagraphic and heliospheric imagery from the STEREO/SECCHI, SOHO/LASCO, and GOES-19/CCOR-1 instruments as the comet approached perihelion while near superior conjunction from Earth. We performed image stacking and photometry with these data, and obtained the following results:

\begin{enumerate}
\item The comet appears extended in a stack of all CCOR-1 frames of the comet, with an apparent ${\sim}4'$ diameter coma.
\item The comet rapidly brightened on its final approach toward perihelion over 2025 September--October at heliocentric distances of $r\lesssim2$~au, with a much steeper $r^{-n}$ brightness scaling of $n=7.5\pm1.0$ than the $n=3.8\pm0.3$ previously reported at $r\gtrsim2$~au. The trend extrapolates to a geocentric $V$ magnitude of $\sim$9 at perihelion.
\item The comet appears distinctly bluer than the Sun in LASCO/CCOR-1 color photometry---in contrast to earlier observations showing the comet's dust to be red---suggesting that gas emission, likely from C$_2$ (and possibly NH$_2$), contributes a sizable fraction of the overall visible brightness.
\end{enumerate}

\begin{acknowledgments}
We thank Quanzhi Ye and Carl A. Schmidt (Boston University) for permitting our use of their spectrum of comet 24P/Schaumasse as a visual aid in Figure~\ref{fig:bands}, Jackie Davies for the HI1 filter transmission data, and Natsuha Kuroda and the CCOR-1 team for the CCOR-1 transmission data. We also thank the reviewer for their suggestions helping to improve the manuscript.

Q.Z. was supported as a Percival Lowell Postdoctoral Fellow at Lowell Observatory. K.B. was supported by the NASA-funded Sungrazer project.

The STEREO/SECCHI data are produced by an international consortium of the NRL, LMSAL, NASA GSFC (USA), RAL and the University of Birmingham (UK), MPS (Germany), CSL (Belgium), and IOTA and IAS (France). The SOHO/LASCO data are produced by a consortium of the NRL (USA), MPS (Germany), Laboratoire d'Astronomie (France), and the University of Birmingham (UK). SOHO is a project of international cooperation between ESA and NASA.
\end{acknowledgments}

\facilities{GOES (CCOR-1), SOHO (LASCO), STEREO (SECCHI)}

\software{\texttt{Astropy} \citep{astropy2013,astropy2018,astropy2022}, \texttt{Astroquery} \citep{ginsburg2019}, \texttt{Matplotlib} \citep{hunter2007}, \texttt{NumPy} \citep{vanderwalt2011,harris2020}, \texttt{SciPy} \citep{virtanen2020}}

\bibliography{ms}{}
\bibliographystyle{aasjournalv7}

\end{document}